\documentclass[11pt,twoside]{article}
\usepackage{cal05}

\contact{Jeremy R. Walsh}
\email{jwalsh@eso.org}

\begin{document}

\title{Recent developments of the ACS spectral extraction software aXe}
\titlemark{ACS spectral extraction software aXe}

\author{Jeremy R. Walsh, Martin K\"{u}mmel, S\o ren S. Larsen}
\affil{Space Telescope European Co-ordinating Facility,
European Southern Observatory, Karl-Schwarzschild-Strasse 2,
D-85748 Garching, Germany}

\paindex{Walsh, J. R.}
\aindex{K\"{u}mmel, M.}
\aindex{Larsen, S. S.}

\authormark{Walsh, K\"{u}mmel \& Larsen}

\begin{abstract}
The software package aXe provides comprehensive spectral extraction  
facilities for all the slitless modes of the ACS, covering the 
Wide Field Channel (WFC) grism, the High Resolution Channel (HRC) 
grism and prism and the Solar Blind Channel (SBC) prisms. The latest 
developments to the package apply to all ACS slitless modes leading to 
improved spectral extraction. Many thousands of spectra may be 
present on a single deep ACS WFC G800L image such that
overlap of spectra is a significant nuisance. Two methods of
estimating the contamination of any given spectrum by its near
neighbours have been developed: one is based on the catalogue of
objects on the direct image; another uses the flux
information on multi-filter direct images. An improvement to the 
extracted spectra can also result from weighted
extraction and the Horne optimal extraction algorithm has
been implemented in aXe. A demonstrated 
improvement in signal-to-noise can be achieved. These new
features are available in aXe-1.5 with the STSDAS 3.4 release. 
\end{abstract}


\section{The spectral extraction package aXe}
As part of a collaborative project between STScI and the Advanced 
Camera for Surveys (ACS) IDT,
the ST-ECF provides comprehensive support for the slitless 
spectroscopy modes of the ACS. As 
well as support to users of the ACS grism (G800L for the Wide Field
Channel, WFC and High Resolution Channel, HRC) and prisms
(PR200L for the HRC and PR110L and PR130L for the Solar Blind Channel,
SBC) and contributions to the ground and in-orbit calibrations of
the slitless modes, a primary pillar of this project has been
the provision of an extraction package, called aXe. The package consists of
a number of self-contained modules, which perform
the basic steps - defining apertures for extraction of a spectrum,
assigning wavelengths to pixels, flat fielding, extracting 
rectified 2D spectra and a 1D spectrum, and applying flux
calibration. The modules are scripted in Python to allow easy 
integration into Pyraf/STSDAS (see K\"{u}mmel et al. 2005
for more details). The aXe user manual 
({\tt http://www.stecf.org/software/aXe/documentation.html})
provides full details for installing and running the package.
 
The fundamental aspect of slitless spectroscopy
is that the individual objects define their own `slit' in terms
of position on the detector and object height of the dispersed
spectrum; the object width in the dispersion direction affects the
spectral resolution. aXe uses a catalogue of the observed targets, 
which is usually taken from a matched direct image
as the starting point of the reduction process. In the design of 
aXe it was decided to make the software as general as possible,
so that in the longer term not only slitless spectra from ACS
could be extracted. This flexibility is engineered by putting
all instrument specific parameters in a configuration file.
Thus the specification of the spectral traces, the dispersion
solutions, the name of the flat field file, the sensitivity file name, 
etc are all listed in a single file for each instrument mode. Thus for
ACS, there are six configuration files; three for the G800L with the
WFC (one for each chip) and HRC, one for PR200L, and one each for 
the SBC with PR110L and PR130L.    
The built-in flexibility has paid off since aXe has also been
used to extract spectra from multi-object spectra taken with 
the VLT FORS2 instrument (K\"{u}mmel et al. 2006).

Since the first release of aXe in 2002 ready for installation
of ACS into HST, the package has evolved. In particular in
2004 a major enhancement was added with the use of `drizzle'
(Fruchter \& Hook, 2002) to combine 2D spectra of individual 
objects when the data is taken with dithers of the telescope.
This turns out to be a very common observational procedure, at 
least for the grism modes, in order to recover the undersampling 
of the Point
Spread Function (PSF) and to mitigate the effect of pixel
sensitivity variations and hot pixels. Since the 2004 release (aXe-1.4),
two enhancements have been added which are 
here described. The sensitivity of the ACS slitless spectroscopy 
modes implies that, despite the small pixel scale and compact PSF,
even for high Galactic latitude fields, the surface density 
of detected spectra on moderate exposures ($>$ thousands of seconds) 
displays crowding and overlap. A high priority was to indicate 
to the user which pixels in an extracted spectrum are affected 
by overlap with spectra of other objects. This has now been implemented
in a quantitative way, whereby the estimated value
of the contaminating flux contributing to a spectrum pixel is
output. The second enhancement was to apply the well known
technique of weighting by the spatial profile when forming 
a 1-D spectrum from the 2D spectrum on the detector (optimal
extraction of Horne 1986). Both these enhancements are described.

\section{Handling spectral contamination in aXe}
The contamination of one spectrum by spectra of neighbouring objects
can be manifest in several ways: 
\begin{itemize}
\item overlap of the first order spectrum by the first order spectrum
of a nearby object situated in the projected direction of
dispersion; 
\item overlap of first order spectra situated nearby but offset perpendicular
to the dispersion direction;
\item both above cases combined and possibly involving spectral orders 
other than the first. 
\end{itemize}
A particular case is that of the zeroth order of a bright object
overlying a fainter object spectrum. For the G800L grism,
the zeroth order is similar in size to the dispersing object, but slightly
dispersed, so that it can resemble a broad emission line. If such a
feature is not
recognized as a contaminating line it may lead to erroneous
wavelength, and hence redshift, assignment. Of course the 
effect of bright objects on faint object spectra is more serious
than the reverse case, and highlights the need for a warning
of the contamination which provides at least an estimate of the
actual contaminating flux to a given spectrum. In the first release
of aXe, a minimal indicator of spectral contamination was 
implemented which indicated the total number of other object spectra 
which fell within a pixel in the extraction box of the given object. With 
aXe-1.5, a new scheme was implemented which provides
a robust estimate of how much contaminating flux contributes to
a given pixel. The contamination is estimated by making a
simulated slitless image, which is achieved by one of two methods.
The simpler method takes the input catalogue and generates simulated
images as 2D Gaussians based on the image parameters, and is called the 
Gaussian Emission Model;
the other method actually uses the fluxes in the pixels of a set
of multi-colour companion direct images, and is called the
Fluxcube Emission Model.

\subsection{Gaussian Emission model}
The input catalogue which drives the object extraction is
usually produced by running SExtractor on a companion direct
image (or several images taken with different filters) and lists
the object position, size parameters and magnitude. Thus for
each object a spectrum over the wavelength extent of the 
slitless dispersing element can be formed and converted to 
detector count rate using the known sensitivity of the slitless 
mode. For a single filter this will be a flat featureless 
spectrum, but more filter bands can more closely match 
the true object spectrum. From the position and extent of each 
object, which are determined from the object parameters of each 
image (A\_IMAGE, B\_IMAGE
and THETA\_IMAGE in SExtractor), a simulated spectrum corresponding to
the slitless spectrum can be formed with spatial extent matching
the object size (see Fig.1, left panel, for an overview). 

\begin{figure}
\epsscale{0.80}
\plottwo{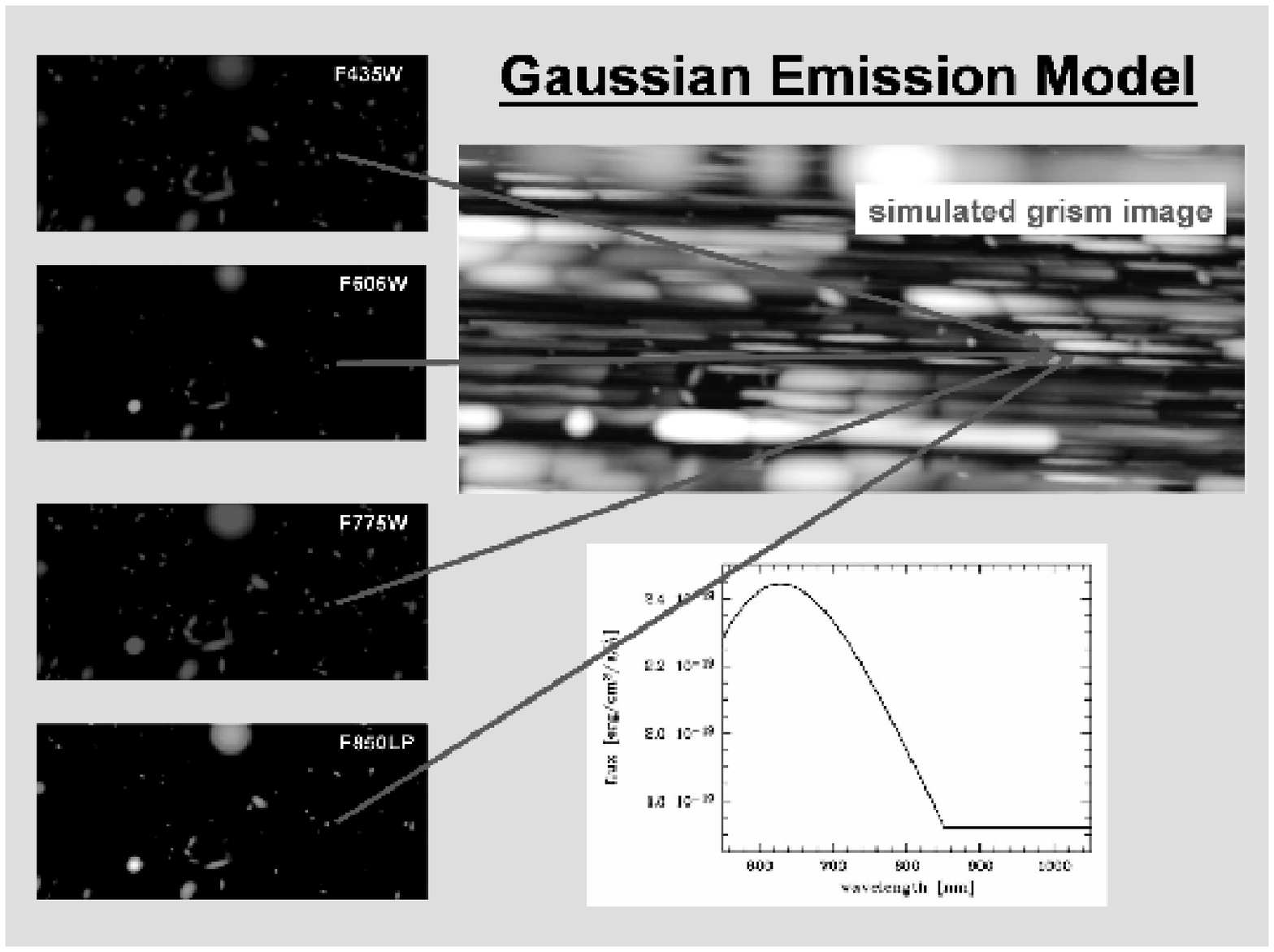}{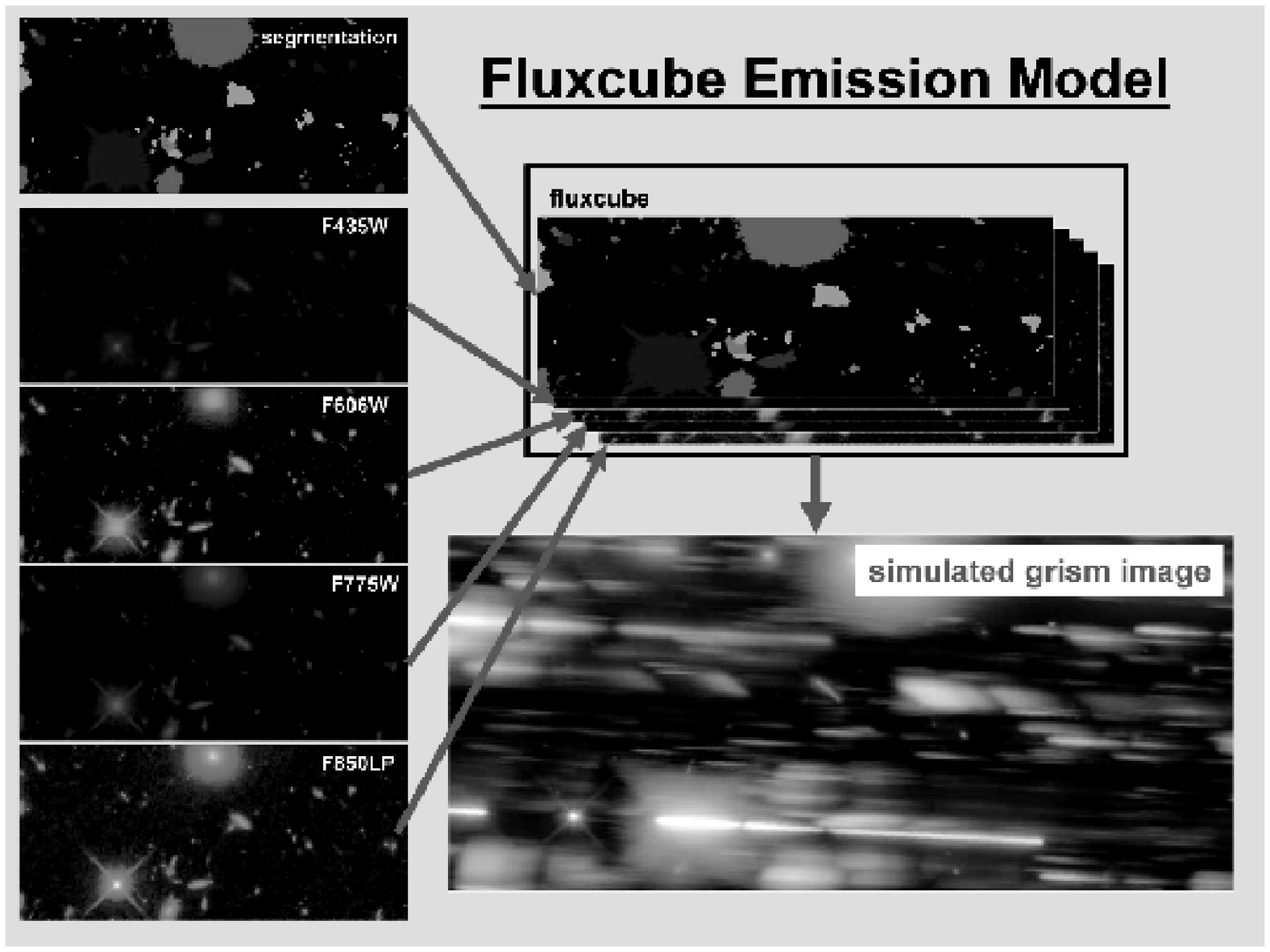}
\caption{ {\bf Left.} The scheme for generating the contamination model based on
2-D Gaussians and an input catalogue with object shapes. The
position, size (major and minor axes and position angle) and magnitude
of objects in different filter direct images provide simulated
objects which are dispersed onto the detector pixels, converted
to $e^{-}$ s$^{-1}$. 
{\bf Right.} As Fig. 1 (left) but showing the procedure for generating the 
simulated images directly from the photometric information in the 
pixels of the multi-colour direct images.}
\label{Fig1}
\end{figure}

\subsection{Fluxcube Emission model}
An alternative, and preferable, method to produce the model
spectrum for estimation of the contribution
of contaminating spectra is to use the surface brightness
distribution in the pixels of the companion direct image(s).
The assignment of pixels to a given object still has to
be established and the method chosen in aXe-1.5 was to use the Segmentation
image provided by SExtractor. This image has the pixel value  
belonging to a given object set to the object number in the
SExtractor catalogue. The data are stored in an intermediate
file which is a cube with planes for the segmentation image
and the filter images in different bands; it is hence called
a Fluxcube (see Fig. 1, right panel). aXe-1.5 provides a 
routine to produce such flux cubes which are then read by the 
appropriate tasks during the extraction of the spectra. 
 
\subsection{Quantitative contamination}
During the extraction process for a given object spectrum,
then, within the extraction box, the count rate in the pixels
belonging to objects other than the one being extracted are
accumulated from the contamination image (originating in
either the Gaussian or Fluxcube models) and assigned to
the contaminating signal. Applying the sensitivity curve to
the contaminating signal enables the total contaminating
signal to the spectrum to be computed. The result
is written to a separate column of the Extracted Spectra File 
by aXe. Fig. 2 shows as an example the extracted spectrum of an 
$i \sim$ 24 mag. emission line galaxy; the strong red continuum
is shown to be badly contaminated (squares show the
contamination contribution). However the pair of emission lines at
around 7500\AA\ can be seen to be intrinsic to the object and not 
arise from a contaminating spectrum.
It must be emphasized that the quantitative contamination is only 
an estimate as it is based on a model (Gaussian or Fluxcube); it 
does however lead to an appropriate level of caution being exercised
in quantitatively assessing a given spectrum. 

The aXe tasks for producing the emission model leading to
contamination estimation deliver simulated images from direct
image(s) or a catalogue based on direct image(s). Thus the
slitless images produced by the Gaussian or Fluxcube models
(....CONT.fits files) are ideally suited to use in observation
planning and exposure time estimation. For example for a complex
field, the contamination images can be used to choose optimal
telescope roll angles that minimize overlap of the spectra of 
interest.

\begin{figure}
\epsscale{0.45}
\plotone{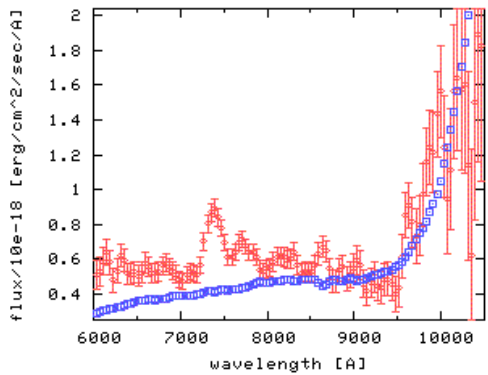}
\caption{The extracted ACS WFC 1D spectrum of an emission line galaxy from
the HUDF parallel data (primary instrument was NICMOS (PI:
Thompson, programme 9803). The flux of the galaxy (with 1$\sigma$ error bars)
(diamonds) and the sum of the contaminating flux in the
extraction box (boxes) are shown.} 
\label{Fig2}
\end{figure}

\section{Weighted spectral extraction}
If the 2D spectrum of an object is extracted applying a set of
weights to the spatial profile, then the resulting
1D spectrum has a higher signal-to-noise than the
simple box-extracted (i.e. summed) spectrum. This
was shown by Horne (1986; see also Robertson 1986) and is
often referred to as optimal extraction. Weighted extraction
has been implemented in aXe-1.5, with the Horne (1986) algorithm, using 
weights derived from the contamination
image described in Section 2. The contamination image is well
suited to providing the weights since it is binned in the same way 
as the spectrum to be extracted and, as a model, does not suffer from 
any systematic statistical effects.
In the examples described by Horne (1986), the weights are
typically determined by fitting the observed spectrum in
the dispersion direction as a function of spatial offset. This
procedure can be prone to failure for weak spectra and where the whole
spectral extent is not occupied by signal; in the case of
overlapping spectra it would provide incorrect weights.

Simulations were performed with a random
star field composed of (spatially) well-sampled star images 
in order to test the optimal extraction implementation in
aXe-1.5. The quantitative contamination procedure in aXe using the 
Gaussian emission model was employed to make simulated slitless 
spectra from a SExtractor catalogue; background and
noise were added and then the spectra were extracted with
and without weighting. The gain of weighted extraction
was specified as the
ratio of the signal-to-noises in the optimal to the
unweighted spectrum over a given range. As the signal-to-noise
level decreases, the optimal extraction shows an advantage
over the unweighted extraction depending on the width of the 
extraction (see Fig. 3 where the
extraction width is 6 times the object width).
At the lowest signal-to-noise level, the advantage is
around a factor 1.4, equivalent to an increase
in exposure time of 1.9 over unweighted extraction. 

As an example of weighted extraction applied to
real ACS data, a crowded
stellar field was selected from archive data resulting from
the APPLES programme
(ACS Pure Parallel Lyman-alpha Emission Survey, PI: J. Rhoads,
Programme 9482). Over 7000 spectra (to $i$=25mag.) were 
present on this
ACS Wide Field Channel image so there is often considerable
contamination between adjacent spectra. Using the
quantitative contamination, only those spectra with $<$5\% 
contamination were selected, and the advantage in 
signal-to-noise of using weighted extraction 
was determined. The mean S/N was computed over a range of
1000\AA\ centred at 7500\AA\ and Fig. 4 shows the result
as a point plot $v.$ magnitude and as a histogram $v.$
signal-to-noise. Here 1374 stars are analysed, and
a peak advantage of weighted over unweighted extraction of
about 1.3 is realised. This is 
somewhat lower than that achieved in the simulations, but
is probably on account of the narrow undersampled Point Spread
Function (PSF) of the WFC data. Nevertheless the theoretical
gain in exposure time is around 60\% at low signal-to-noise.

Weighted extraction can be selected 
in the aXe parameter set and both weighted and unweighted 1D spectra
for all sources are output. For sources with complex cross
dispersion profiles, there will probably be little advantage 
of weighted extraction but for small objects, such as faint stars
and distant galaxies, a modest gain in signal-to-noise is 
achievable with optimal extraction for slitless spectra. \\

The latest release aXe-1.5 provides both quantitative contamination
and weighted extraction for all ACS slitless spectroscopy modes and 
is available from November 2005, with Pyraf/STSDAS 3.4. Full details 
can be found at \\
{\tt http://www.stecf.org/software/aXe/}

\begin{figure}
\epsscale{0.42}
\plotone{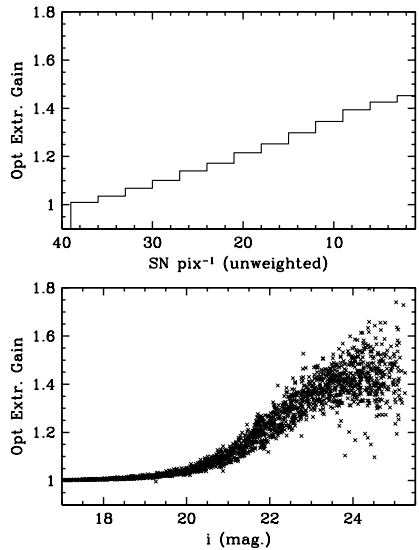}
\caption{The result of a simulation of a star field observed with 
the ACS
WFC G800L grism in terms of the signal-to-noise (S/N) advantage 
of the weighted over the unweighted extraction (extraction 
widths $\pm$3 $\sigma$ of the sources). The lower plot shows 
the actual advantage for each star in the simulation (ratio
of weighted S/N divided by unweighted S/N, averaged over a 
wavelength range) in terms of the $i$ mag. 
The upper plot shows a histogram version of the lower plot where 
the plot abscissa is the mean signal-to-noise over a 1000\AA\ 
range.
} 
\label{Fig3}
\end{figure}

\begin{figure}
\epsscale{0.42}
\plotone{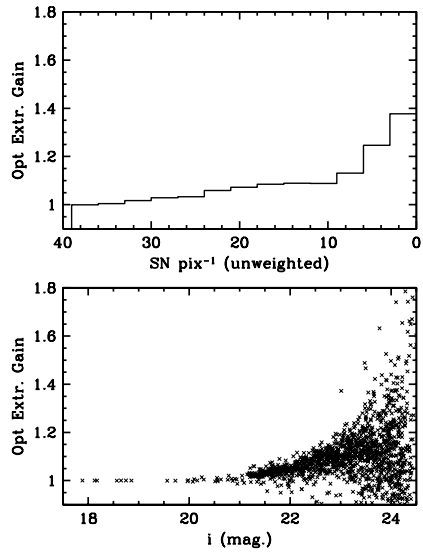}
\caption{
An identical plot to Figure 3, but now based on real data taken 
with the ACS WFC G800L grism as part of HST APPLES programme 9482.
The field is at low Galactic
latitude and consists of many late type stars. There is considerable
scatter in the weighted v. unweighted extraction advantage at the lowest signal
level, but a demonstrable increase in signal-to-noise at low
flux levels is achieved.
} 
\label{Fig4}
\end{figure}

\end{document}